\documentclass[conference]{IEEEtran}
\IEEEoverridecommandlockouts
\usepackage{cite}
\usepackage{amsmath,amssymb,amsfonts}
\usepackage{algorithmic}
\usepackage{graphicx}
\usepackage{textcomp}
\usepackage{xcolor}
\usepackage{booktabs}
\usepackage{lipsum}
\usepackage{soul}
\usepackage{caption}
\usepackage{subcaption}
\usepackage{siunitx}
\usepackage{url}

\def\BibTeX{{\rm B\kern-.05em{\sc i\kern-.025em b}\kern-.08em
    T\kern-.1667em\lower.7ex\hbox{E}\kern-.125emX}}

\begin{document}

\title{Comparative Study of Hardware and Software Power Measurements in Video Compression
{
\thanks{This work has been supported by the UKRI Creative Bristol+Bath Cluster, the UKRI MyWorld Strength in Places Programme (SIPF00006/1), and the Bristol Digital Futures Institute.}
}}

\author{Angeliki Katsenou, Xinyi Wang, Daniel Schien, and David Bull\\
Visual Information Lab, School of Computer Science, University of Bristol, Bristol BS1 8UB, UK\\
\{firstname.lastname\}@bristol.ac.uk }

\maketitle

\begin{abstract}
The environmental impact of video streaming services has been discussed as part of the strategies towards sustainable information and communication technologies. A first step towards that is the energy profiling and assessment of energy consumption of existing video technologies. This paper presents a comprehensive study of power measurement techniques in video compression, comparing the use of hardware and software power meters. An experimental methodology to ensure reliability of measurements is introduced. Key findings demonstrate the high correlation of hardware and software based energy measurements for two video codecs across different spatial and temporal resolutions at a lower computational overhead.
\end{abstract}


\section{Introduction}
The proliferation of video streaming services in recent years has undeniably revolutionised the way we consume entertainment~\cite{sandvine}, but it has also raised significant concerns about its environmental impact. The surge in online video content, from streaming platforms to user-generated content, has led to immense data centre deployment, network traffic, and user data consumption, resulting in high energy consumption\cite{CarbonTrust2021}. The massive data centres and networks required to support these services rely heavily on electricity, a significant portion of which is still generated from fossil fuels. This, in turn, contributes to increased carbon dioxide (CO2) emissions, exacerbating the challenges of global climate change, as identified in the Paris agreement in 2015~\cite{Greenpeace2015}. 

The environmental impact of video streaming services goes beyond just data centres and network infrastructure. User devices play a crucial role in exacerbating the problem. With the vast and ever-expanding audiences these services attract, individual consumption is scaled up significantly~\cite{sandvine}. Smartphones, tablets, laptops, TVs, etc, which are used to access and view video content, require substantial amounts of electricity to operate, which is produced by fossil fuels~\cite{CarbonTrust2021}. Furthermore, as new media formats emerge and appetite for more immersive experiences grows (higher spatio-temporal resolutions, volumetric video, etc), the energy demands of the new media delivery devices are expected to grow as well. 
Hence, it becomes imperative for both service providers and consumers to design more energy-efficient technologies and practices to minimize the environmental consequences of our digital entertainment choices. 

The energy profiling of video services depends on the reliability of power measurements. The methodologies around power measurements vary with the application. There are two experimental approaches: the use of hardware (HW) power meters and the use of software (SW) based power meters. The former are devices that measure the electricity intensity at the power supply, while the latter are sampling the usage of internal hardware, e.g., CPU, and typically use thermal design power based modeling.
The hardware power meters are less intrusive, however, do not detail the consumption of the various components or services launched on the computing node. On the other hand, the software power meters use internal interface directly to retrieve power consumption of the underlying hardware being used. Therefore depending on the application/service to be measured an exploration of the feature set (e.g., sampling rate) and performance of hardware or software-based power meters should be tested~\cite{RAPL2018}. Khan et al~\cite{RAPL2018} performed a comparison of using Running Average Power Limit (RAPL) or a smart power plug to measure the energy consumption for large scale experiments on cloud computing servers and concluded that RAPL is reliable. Another interesting benchmark study on software based power measurements is reported by Jay et al~\cite{Jay_IEEEACMCCGrid}. This work concluded that there is a significant but not constant offset between the external hardware power meters and the software-based power meters. 
In the field of video streaming, some of works on monitoring the energy consumption were based on external hardware power meters~\cite{Li_VCIP2012, Noureddine2013, Preist_CHI2019, Mercat_ICIP2023}, while others employed software-based power meters, commonly tools built on RAPL~\cite{Monteiro_ISCAS2015, KatsenouPCS2022, HerglotzCSVT2019, KraenzlerPCS2022, Amripour_ICME2023, Chachou_MMSP2023}, such as the Intel Power Gadget~\cite{IntelPowerGadget} or the CodeCarbon tool~\cite{CodeCarbon}. The SW power meters have the advantage of low cost and the direct measurement of power required by particular processes instead of a whole workstation.

In this work, in an effort to verify the validity of measurements based on software power meters, we perform a comparative study of hardware against software power meters. This study contributes to the field by providing a nuanced understanding of power measurement methodologies, vital for advancing energy-efficient video processing technologies. Taking into consideration that video encoding and decoding are complex processes with different performance across different content and video parameters, we introduce an experimental pipeline to ensure statistical validity of the measurements. For the experiments, we consider a selection of various genres of native 4K video sequences at different frame rates. To complement this study, we included different spatial resolutions by downsampling the video sequences to lower spatial resolutions. Regarding the codecs, we consider H.264/MPEG-4-AVC~\cite{r:h264} and H.265/HEVC~\cite{r:HEVC} as two of the most deployed video technologies. 

The remainder of this paper is organized as follows. Section~\ref{sec: MeasurementMethod} explains the different power measurements and the computation of energy. Section~\ref{sec: Experiments} details the experimental set up, while results are reported in Section~\ref{sec: Results}. Finally, conclusions are drawn in Section~\ref{sec: Conclusion}.

\section{Measuring Energy}
\label{sec: MeasurementMethod}
In this work, the aim is to measure the energy consumption on both the encoder and the decoder ends. The encoding side is a good representation of the energy consumption at the video provider's side (e.g., in data centers), while the decoding concerns the part of the energy consumed at the end-user devices (only for the decoding, not the display). We first measure the power consumed at each sampling interval and then compute the energy over an observed time interval as the integral of power. Particularly, the encoding energy can be obtained by:
\begin{equation}
    E_{enc}=\int^{T_{enc}}_{t=0} P_{enc}(t) dt ,
\end{equation}
where $T_{enc}$ is the encoding duration. The same principle applies for any other energy measurement, such as decoding or measurement of idle energy (when no process is executed in order to measure a base energy consumption of background processes).

In order to ensure that any increases in power consumption were specifically attributed to the encoding and decoding processes, the encoding server and decoding machine were used exclusively for the experiment. However, it is important to note that there was some unavoidable background power consumption noise and uncertainty due to the operating system's background activities.

\subsection{Software-based Power Measurements}
Our performance investigation for software-based power measurements relies on the integrated power meter in Intel CPUs, the RAPL~\cite{RAPL2018}. 
RAPL reports the energy consumption on different levels or power domains: entire CPU socket (PKG), all CPU cores (PP0), integrated graphics (PP1), and dynamic random-access memory (DRAM). The availability of power domains may vary between architectures and processor models. The RAPL energy calculations are implemented in the hardware, so energy consumption can be measured without significant computational overhead. No additional equipment is needed, which makes it a cost-effective alternative to HW power meters~\cite{RAPL2018}. A minor drawback of RAPL is that it does not provide timestamps, but rather a relative timing of each sample from the start time of the task execution (set to 0).
Tools based on RAPL have been used in other similar research activities (e.g.,~\cite{KaupTCSVT2016, KaupVCIP2020, KatsenouPCS2022, Chachou_MMSP2023}).


\begin{figure}[!t]
	\centering
		\includegraphics[scale=.32]{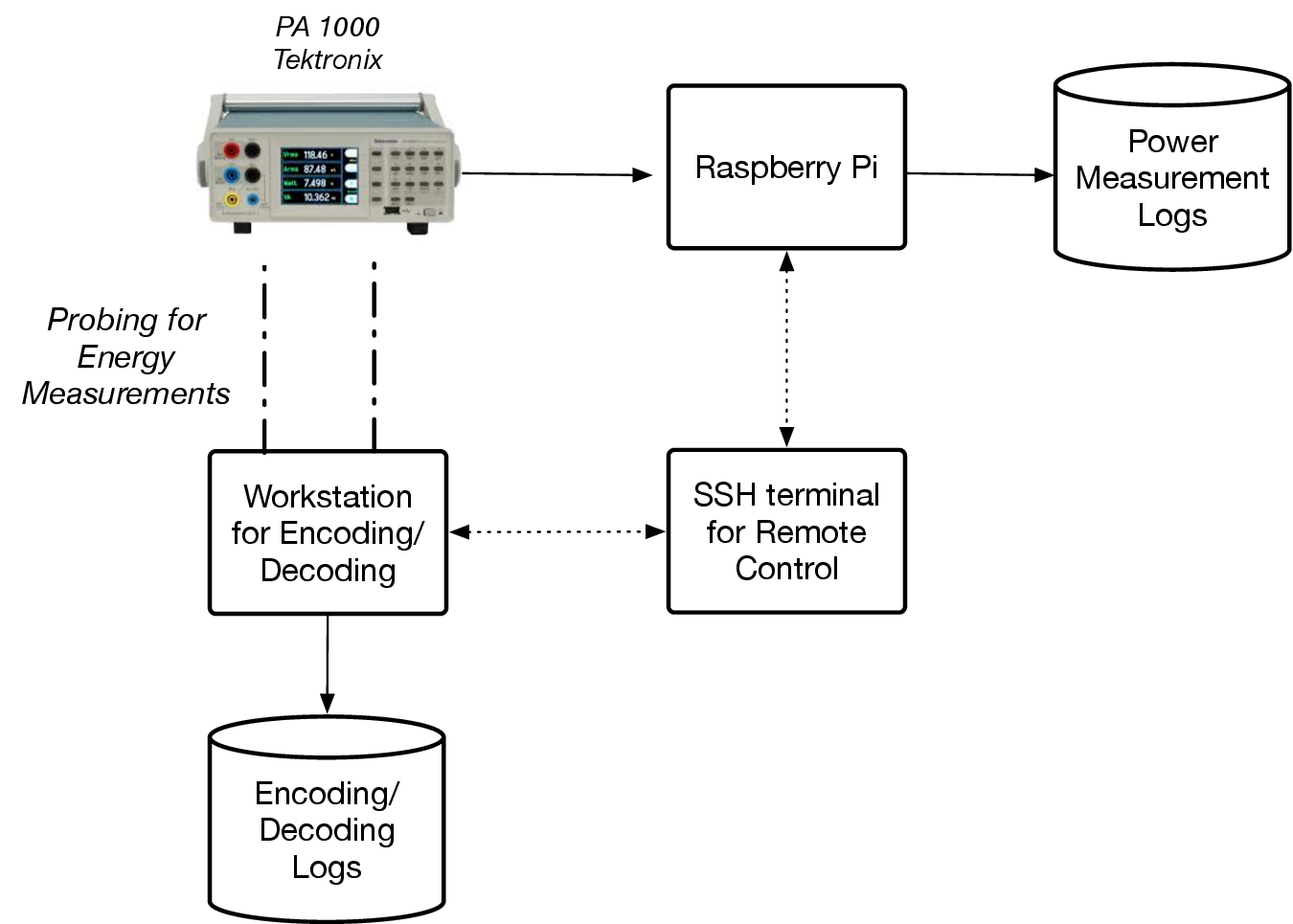}
	\caption{Experimental topology for measurements using a HW power meter. 
 }
	\label{fig:TestSetup}
\end{figure}

\begin{figure*}
	\centering
		\includegraphics[scale=.32]{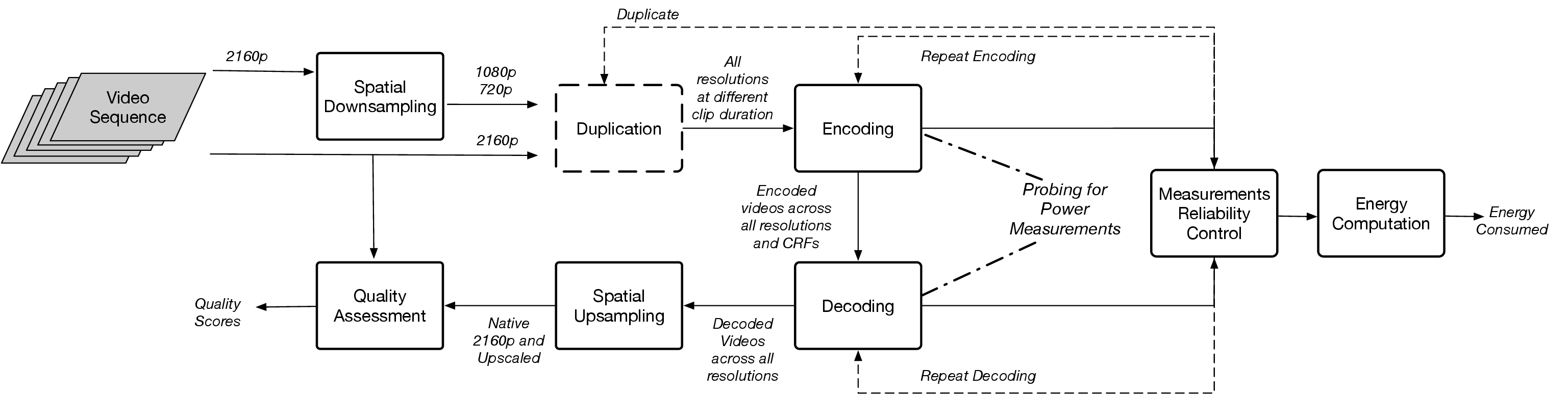}
	\caption{Experimental pipeline for measuring the encoding and decoding energy across different spatial resolutions either with hardware or software power meters. The dashed lines/outlines indicate conditional processes based on the reliability control, while he dash-dotted lines the probes for the power measurements.}
	\label{fig:Overview}
\end{figure*}

\subsection{Hardware-based Power Measurements}
A HW power meter is an external device that needs to be connected to the workstation utilised for encoding/decoding. The experimental setup we used for measuring energy consumption in this case is depicted in Fig.\ref{fig:TestSetup}. In this figure, a Tektronix PA1000 power meter is connected~\cite{HWPowermeter} to measure the energy consumed on the workstation. The codec logs are stored in a database in the workstation. A Raspberry Pi is used to log the power measurements to another database. The logs from the power measurements and encoding/decoding are accessed through SSH via another computing node after the end of the measurements to eliminate any additional power consumption noise caused by the display controller.
 Throughout the experiments, power consumption was measured at a half-second interval for both encoding and decoding.

The timestamps tagging the codec logs and power measurements in the different drives, which are depicted as being stored in two separate databases in Fig.~\ref{fig:TestSetup}, are based on the network time protocol. As a result, the timestamps in the databases are accurate, but need alignment. Therefore, for each start and end timestamps in the job logs database, the nearest timestamps from the power measurements database were identified. These aligned timestamps are then used to calculate the energy consumed for encoding, $E^{H}_{enc}$. Similarly, the timestamps of the decoding logs need to be aligned in order to compute the decoding energy $E^{H}_{dec}$. 

The encoding energy measure with the HW power meter includes energy consumed on the whole workstation and can be roughly expressed as the aggregation of energy spent at the processor $E_{proc}$, storage $E_{strg}$, and energy used for other background processes, including cooling $E_{x}$, i.e.:
\begin{equation}
    E^H_{enc} = E_{proc} + E_{strg} + E_{x} .
    \label{eq:ODC}
\end{equation}
\noindent
In a streaming scenario, the decoded video is not stored on a local drive. Therefore, the decoding energy comprises of:
\begin{equation}
    E^H_{dec} = E_{proc} + E_{x} .
    \label{eq:User}
\end{equation}

\subsection{Reliability of Power Measurements}
\label{ssec: Reliability}
A major factor of the measurements and their reliability is the sampling rate. RAPL has a default sampling rate of 10Hz (1 sample every 100ms), which we used for this work. The Tektronix PA1000~\cite{HWPowermeter} has a lower sampling rate of 2Hz (one sample every 500ms). The sampling rate is important for the reliability and statistical validity of the measurements. For processes of short execution time, the lower the sampling rate, the harder it is to measure reliably the power consumption from a small set of samples. This is particularly important for decoding, as for most of the cases the decoding time is very short (usually less than 1 second per 60 frames). Thus, the calculation of the decoding energy is sensitive to power noise caused by the operating system's background activities.
Therefore, we introduced a few techniques that allow us acquiring statistically reliable measurements. The first was to duplicate the video sequences from 20sec into longer versions to actively increase the encoding and decoding time. The decoding time was more critical as it is the fastest process whose energy consumption needs to be captured with validity. The second technique was to repeat the same encoding/decoding process until the number of acquired power samples was providing statistically reliable measurements. Similarly to~\cite{Ramasubbu_PCS2022}, we test the distribution of our measurements, as follows:
\begin{equation}
    \dfrac{t_{\frac{\alpha}{2}}\sigma}{2\alpha \overline{P}} < N \sqrt{2N} \, ,
\end{equation}
where $N$ is the number of samples, $\overline{P}$ is the mean power, $\sigma$ is the standard deviation of the power distribution, $t_{\frac{\alpha}{2}}$ is the critical $t$-value of Student's $t$ distribution, and $\alpha$ is the confidence bound. 

The number of duplications is mainly determined by the time duration of the encoding or decoding task which depends on the spatial and temporal resolution as well as on the codec. For example, a 1080p sequence of \SI{15}{fps} is encoded and decoded faster than one of \SI{60}{fps}. 

\section{Experimental Design}
\label{sec: Experiments}
In this section, we provide in detail the experimental design and parameters selected. An overview of the experimental pipeline is illustrated in Fig.\ref{fig:Overview}. The 2160p native video sequences are first downscaled to 1080p and 720p using the three-tap Lanczos~\cite{Duchon} filter. Next to that, the sequences are duplicated multiple times depending on their specs and method. For example, for software-based measurements only the videos of 720p require a duplication. After duplication, the actual encoding at different compression levels and then decoding is executed. These are the two processes that are probed separately to be measured for the power consumption and energy calculation either by the use of a software-based or hardware-based power meters. After decoding the 1080p and 720p sequences, are upscaled to 2160p for quality metrics computation. For this study, we do not include the energy consumption from display. The workstation used for the compression experiments has an Intel(R) Core(TM) i9-7900X CPU @3.30GHz and 64GB RAM.

\subsection{Test Video Sequences}
We selected sequences of the YouTube-UGC~\cite{AdsumilliMMSP2019} dataset, as it comprises different genres and is representative of streamed user generated content. 
13 2160p native sequences from the Animation, Gaming, Sports, HDR, and Vlogs genres were randomly chosen. All videos are of 20sec duration with a YUV 4:2:0 color sampling, and frame rates ranging from 15 to \SI{60}{fps} at a native spatial resolution of 2160p\footnote{A full list of the sequences used along with the power measurements is available on the project page~\cite{ProjectPage_Xinyi}.}.

\subsection{Video Codecs}
For this study, we used the ffmpeg N-110021-g85b185b504 version~\cite{ffmpeg} implementations of H.264/MPEG-4-AVC~\cite{r:h264} and H.265/HEVC~\cite{r:HEVC, j:Ohm, b:Wien}. H.264/MPEG-4-AVC was launched in 2004 and remains one of the most widely deployed video coding standards, even though the next generations of standards. H.265/HEVC provides enhanced encoding performance compared to its predecessor. We used the default presets (medium) in the Constant Rate Factor (CRF) mode, which allows consistent quality across frames. In Table~\ref{tab:codecs}, we provide the commands used for the encoding and decoding, as well as the values of the parameters.

\begin{table} 
\caption{The encoding/decoding commands and parameters.}
\centering 
\begin{tabular}{p{2.4cm} | p{5.9cm} }
\toprule
Process/Parameter & Command Invocations \\
\midrule
Encoding &\texttt{ffmpeg.exe -s \$widthx\$height   -r \$FPS -pix\_fmt \$YUVfmt -i \$input.yuv -c:v \$codec -crf \$QP \$encoded.mp4 } \\
\midrule
Decoding &\texttt{ffmpeg.exe -i \$encoded.mp4 -f null } \\
\midrule
\texttt{\$codec} & \texttt{libx264} or \texttt{libx265}\\
\midrule
\texttt{\$crf} & $\{10,20,\ldots,50\}$\\
\midrule
\texttt{\$widthx\$height} & $\{3840\times2160, 1920\times1080, 1280\times720\}$\\
\midrule
\texttt{\$FPS} & $\{15,24,30,60\}$\\
\midrule
\texttt{\$YUVfmt} & \texttt{yuv420}\\
\bottomrule
\end{tabular} 
\label{tab:codecs}
 \vspace{-2em}
\end{table}

\section{Results and Discussion}
\label{sec: Results}

\begin{figure}
    \begin{minipage}[b]{\linewidth}
        \centering
        \includegraphics[scale=.45]{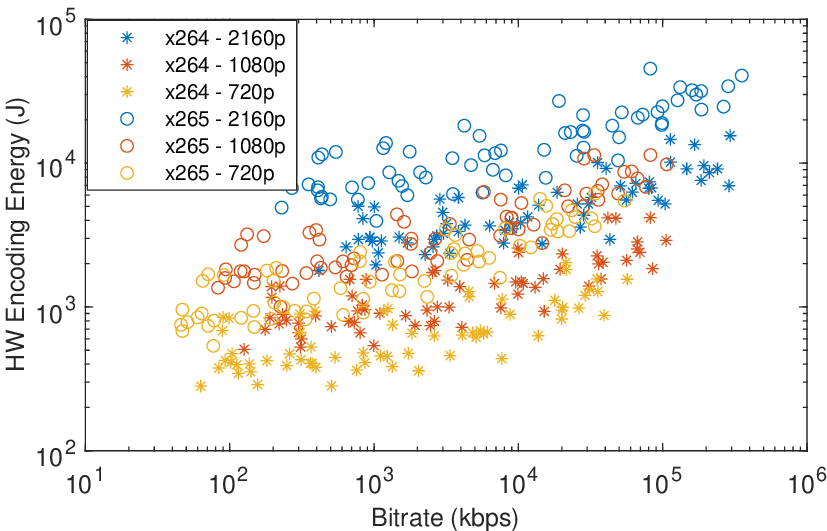}
        \subcaption{HW measurements.}
        \vspace{.1em}
    \end{minipage}
    \begin{minipage}[b]{\linewidth}
        \centering
        \includegraphics[scale=.45]{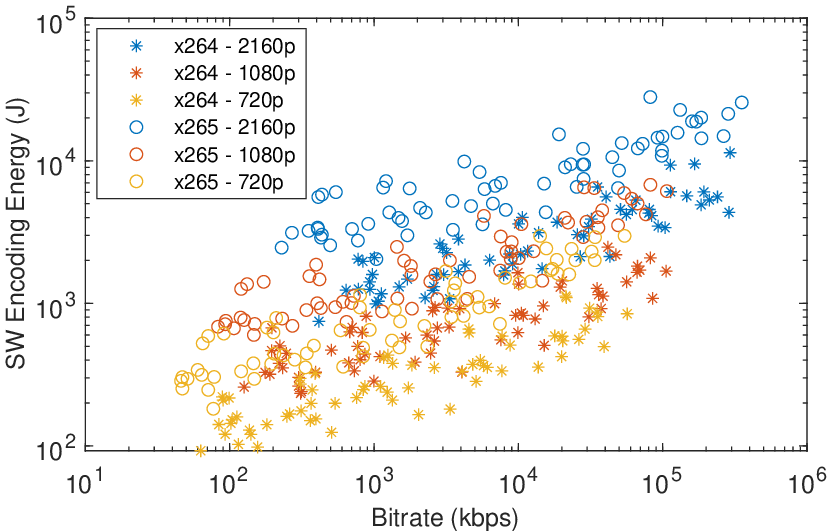}
        \subcaption{SW measurements.}
        \vspace{.1em}
    \end{minipage}
    \begin{minipage}[b]{\linewidth}
        \centering
        \includegraphics[scale=.45]{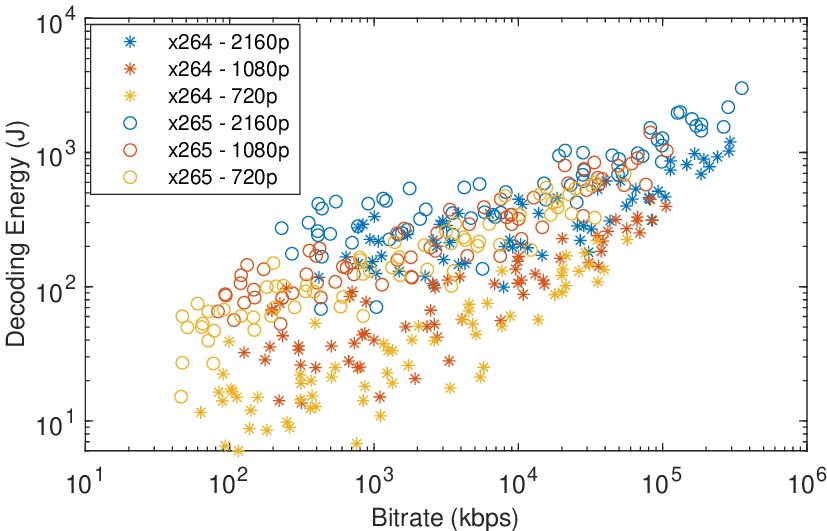}
        \subcaption{HW measurements.}
        \vspace{.1em}
    \end{minipage}
    \begin{minipage}[b]{\linewidth}
        \centering
        \includegraphics[scale=.45]{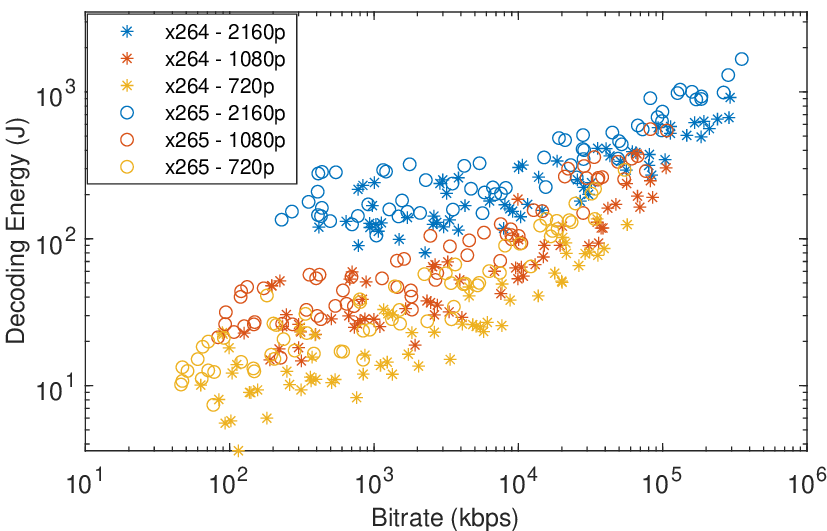}
        \subcaption{SW measurements.}
        \vspace{.1em}
    \end{minipage}
    \caption{Scatter plots of encoding  and decoding energy against bitrate for both codecs. The axes are on logarithmic scales for better illustration of the whole range of measurements.}
\label{fig: EnergyAcrossRes}
 \vspace{-2em}
\end{figure}

A first visual inspection of the collected measurements with the two different methods can be made through Fig.~\ref{fig: EnergyAcrossRes}, where the SW and HW based computations of encoding and decoding energy are scattered across the bitrates of the encoded videos. It is easily observed that the plane of measurements looks very similar between HW and SW methods, just shifted, indicating high correlation. Furthermore, patterns that follow the resolution groups or codecs can also be clearly observed. For example, the cloud of x.265 points appears shifted to the left, i.e., higher compression rate, and upwards, i.e., higher energy consumption, compared to x.264 for all resolutions.

Given that HW power meters measure the consumption by the whole workstation, while RAPL only the one of the chip, we could not perform a direct one-to-one numerical comparison. Therefore, Pearson (PCC), Spearman (SCC), and Kendall (KCC) cross-correlation values were computed between the two approaches for encoding and decoding energy measurements for both codecs. These values are reported in Table~\ref{tab: corrMetrics} and reveal a high linear and rank correlation between HW and SW measurements. SCC and KCC values that reflect the rank correlation are high showing that the relationship is monotonic. We notice, however, slightly reduced KCC values compared to the others. KCC captures the discordant measurements, that could potentially be considered as outliers. These are probable during decoding where the magnitude order of the power consumed could be potentially affected by background processes. Nevertheless, the number of these measurements is not high enough to distort the main conclusions. 

\begin{figure}[h]
    \begin{minipage}[b]{.47\linewidth}
        \centering
        \includegraphics[scale=.5]{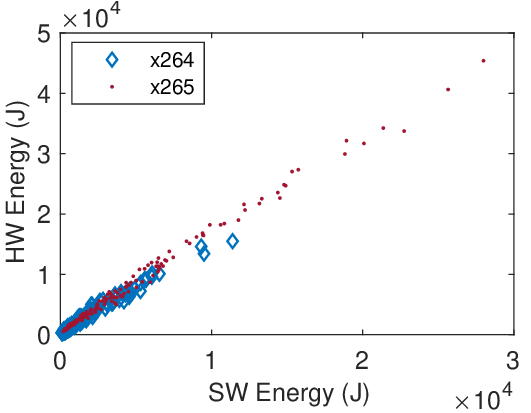}
        \subcaption{Encoding Energy.}
         \vspace{-.5em}
    \end{minipage}
\hfill
     \begin{minipage}[b]{.47\linewidth}
        \centering
        \includegraphics[scale=.5]{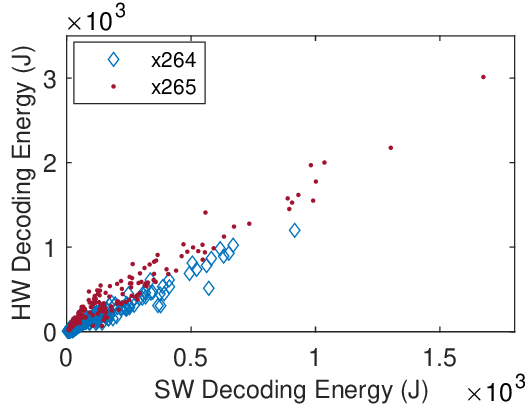}
        \subcaption{Decoding Energy.}
         \vspace{-.5em}
    \end{minipage}
    \caption{Scatter plots of encoding and decoding energy using SW and HW power meters for both codecs. 
    }
\label{fig: EncDec_Scatterplots}
\vspace{-1.2em}
\end{figure}

To further complement the cross-correlation values, Fig.~\ref{fig: EncDec_Scatterplots} illustrates the linear correlation of the HW and SW based measurements for both video codecs through scatter plots. From these plots, it is also clearly observed that the encoding energy measurements have a tighter distribution compared to the decoding measurements that are more prone to noise from background processes. Moreover, we fit linear models that confirmed based on the SW measurements, we can easily estimate the respective HW measurements through the fitted linear models that demonstrate high R$^2$ values ($\ge0.95 $). These results are confirmed by the mean relative estimation error $\epsilon$ of the HW energy consumption~\cite{Ramasubbu_PCS2022}. 


Another aspect for comparison between the HW and SW measurements for video coding tasks is related to the computational overhead that each method brings. As explained in Section~\ref{ssec: Reliability}, the sampling rate is key to ensure statistical significance of the measurements. As the Tektronix HW power meter used in this case has a sampling rate of approximately 1 sample per \SI{500}{ms}, we expect that a higher number of iterations of encoding or decoding should be necessary for reliable measurements. For this set of experiments, the average number of single encodes/decodes for x.264 was 8.67/10.87 and for x.265 8.67/8.68, respectively. These numbers drop down to an average of 1.33/1.33 for both codecs in the case of RAPL due to its higher sampling rate that can be adjusted up to 1 sample per \SI{1}{ms}.

Overall, through this study, we safely conclude that using RAPL, a SW power meter that is easy to install and run, we can reliably measure the codec energy consumption at a chip level with less overhead compared to the HW power meter we compared to. This can provide a safe basis for the design and assessment of the energy profile of new video  technologies.

\begin{table} [h]
\caption{The correlation values of HW vs SW encoding/decoding measurements for both codecs and the goodness of fit metrics, R$^2$ and $\epsilon$.
}
\centering 
\begin{tabular}{l|r|r|r|r|r}
\toprule
Codec - Process &  PCC($\uparrow$) & SCC($\uparrow$) & KCC($\uparrow$)& R$^2$($\uparrow$)& $\epsilon$ ($\downarrow$)\\
\midrule
x264 - Encoding & 0.99&0.98 & 0.90 & 0.97& 11.05\%\\
\midrule
x264 - Decoding & 0.98& 0.97&0.86 & 0.95& 20.05\%\\
\midrule
x265 - Encoding& 0.99&0.99 &0.95 & 0.99& 6.40\%\\
\midrule
x265 - Decoding& 0.98& 0.94& 0.82 & 0.95 & 18.77\%\\
\bottomrule
\end{tabular} 
\label{tab: corrMetrics}
 \vspace{-.5em}
\end{table} 

\section{Conclusion}
\label{sec: Conclusion}
In this paper, we performed a comparative study of acquiring power measurements for video codecs either using an external HW device on the workstation or a software power meter that samples the power consumed on the processing unit. The study concludes that these methods result in linearly correlated measurements for encoding and decoding energy consumption. Further to this, we fitted linear models that further shown the effectiveness of SW meters in approximating HW meter readings. Moreover, the SW power meters bring the benefit of reduced computational overhead. This demonstrates their importance in energy profiling of video services.
This finding is crucial for designing SW-based energy assessment methods that can lead to effective interventions. Hence, part of our future work is to design light-weight energy estimation models that will be deployed for the optimisation of the video streaming pipeline.

\bibliographystyle{ieeetr}
\bibliography{refs}

\begin{thebibliography}{10}

\bibitem{sandvine}
Sanvdine, ``The global internet phenomena report january 2023.''
  \url{https://www.sandvine.com/global-internet-phenomena-report-2023-download?submissionGuid=bd3de666-249c-4927-9bab-b84c3577b2c9}.
\newblock [Online; accessed 20-Mar-2023].

\bibitem{CarbonTrust2021}
{Carbon Trust}, ``{Carbon impact of video streaming},'' tech. rep., 2021.

\bibitem{Greenpeace2015}
Greenpeace, ``{Clicking Clean: A Guide to Building the Green Internet},'' 2015.

\bibitem{RAPL2018}
K.~N. Khan, M.~Hirki, T.~Niemi, J.~K. Nurminen, and Z.~Ou, ``{RAPL in Action:
  Experiences in Using RAPL for Power Measurements},'' {\em ACM Trans. Model.
  Perform. Eval. Comput. Syst.}, vol.~3, no.~2, 2018.

\bibitem{Jay_IEEEACMCCGrid}
M.~Jay, V.~Ostapenco, L.~Lefevre, D.~Trystram, A.-C. Orgerie, and B.~Fichel,
  ``An experimental comparison of software-based power meters: focus on cpu and
  gpu,'' in {\em 2023 IEEE/ACM 23rd International Symposium on Cluster, Cloud
  and Internet Computing (CCGrid)}, pp.~106--118, 2023.

\bibitem{Li_VCIP2012}
X.~Li, Z.~Ma, and F.~C.~A. Fernandes, ``Modeling power consumption for video
  decoding on mobile platform and its application to power-rate constrained
  streaming,'' in {\em 2012 Visual Communications and Image Processing},
  pp.~1--6, 2012.

\bibitem{Noureddine2013}
A.~Noureddine, R.~Rouvoy, and L.~Seinturier, ``A review of energy measurement
  approaches,'' {\em SIGOPS Oper. Syst. Rev.}, vol.~47, no.~3, p.~42–49,
  2013.

\bibitem{Preist_CHI2019}
C.~Preist, D.~Schien, and P.~Shabajee, ``Evaluating sustainable interaction
  design of digital services: The case of youtube,'' in {\em CHI Conference on
  Human Factors in Computing Systems}, p.~1–12, Association for Computing
  Machinery, 2019.

\bibitem{Mercat_ICIP2023}
P.~Sjövall, A.~Mercat, and J.~Vanne, ``Fpga-accelerated hevc encoder for
  energy-efficient multi-access edge computing,'' in {\em IEEE International
  Conference on Image Processing}, pp.~2215--2219, 2023.

\bibitem{Monteiro_ISCAS2015}
E.~Monteiro, M.~Grellert, S.~Bampi, and B.~Zatt, ``Rate-distortion and energy
  performance of hevc and h.264/avc encoders: A comparative analysis,'' in {\em
  IEEE International Symposium on Circuits and Systems}, pp.~1278--1281, 2015.

\bibitem{KatsenouPCS2022}
A.~Katsenou, J.~Mao, and I.~Mavromatis, ``Energy-rate-quality tradeoffs of
  state-of-the-art video codecs,'' in {\em Picture Coding Symposium (PCS)},
  pp.~265--269, 2022.

\bibitem{HerglotzCSVT2019}
C.~Herglotz, A.~Heindel, and A.~Kaup, ``Decoding-energy-rate-distortion
  optimization for video coding,'' {\em IEEE Transactions on Circuits and
  Systems for Video Technology}, vol.~29, no.~1, pp.~171--182, 2019.

\bibitem{KraenzlerPCS2022}
M.~Kr\"anzler, A.~Kaup, and C.~Herglotz, ``Advanced design space exploration
  for joint energy and quality optimization for vvc,'' in {\em Picture Coding
  Symposium (PCS)}, 2022.

\bibitem{Amripour_ICME2023}
H.~Amirpour, V.~V. Menon, S.~Afzal, R.~Prodan, and C.~Timmerer, ``Optimizing
  video streaming for sustainability and quality: The role of preset selection
  in per-title encoding,'' in {\em IEEE International Conference on Multimedia
  and Expo}, pp.~1679--1684, 2023.

\bibitem{Chachou_MMSP2023}
T.~Chachou, W.~Hamidouche, S.~A. Fezza, and G.~Belalem, ``Energy consumption
  and carbon footprint of modern video decoding software,'' in {\em IEEE 25th
  International Workshop on Multimedia Signal Processing}, pp.~1--6, 2023.

\bibitem{IntelPowerGadget}
``{Intel Power Gadget}.''
  \url{https://www.intel.com/content/www/us/en/developer/articles/tool/power-gadget.html}.

\bibitem{CodeCarbon}
``{Code Carbon}.'' \url{https://mlco2.github.io/codecarbon/index.html}.

\bibitem{r:h264}
{ITU-T Rec. H.264}, ``Advanced video coding for generic audiovisual services,''
  2005.

\bibitem{r:HEVC}
{ITU-T Rec H.265}, ``High efficiency video coding,'' 2015.

\bibitem{KaupTCSVT2016}
C.~{Herglotz}, D.~{Springer}, M.~{Reichenbach}, B.~{Stabernack}, and A.~{Kaup},
  ``{Modeling the Energy Consumption of the HEVC Decoding Process},'' {\em IEEE
  Transactions on Circuits and Systems for Video Technology}, vol.~28, no.~1,
  pp.~217--229, 2018.

\bibitem{KaupVCIP2020}
M.~{Kr{\"a}nzler}, C.~{Herglotz}, and A.~{Kaup}, ``{DENESTO: A Tool for Video
  Decoding Energy Estimation and Visualization},'' in {\em Proc. of 2020 IEEE
  International Conference on Visual Communications and Image Processing
  (VCIP)}, pp.~259--259, 2020.

\bibitem{HWPowermeter}
``{Tektronix Power}.'' \url{https://github.com/sust-cs-uob/tek_power/}.

\bibitem{Ramasubbu_PCS2022}
G.~Ramasubbu, A.~Kaup, and C.~Herglotz, ``A bit stream feature-based energy
  estimator for {HEVC} software encoding,'' in {\em Picture Coding Symposium},
  pp.~19--23, 2022.

\bibitem{Duchon}
C.~E. Duchon, ``{Lanczos filtering in one and two dimensions},'' {\em Journal
  of Applied Meteorology}, vol.~18, no.~8, pp.~1016--1022, 1979.

\bibitem{AdsumilliMMSP2019}
Y.~Wang, S.~Inguva, and B.~Adsumilli, ``{YouTube UGC Dataset for Video
  Compression Research},'' in {\em IEEE 21st International Workshop on
  Multimedia Signal Processing (MMSP)}, 2019.

\bibitem{ProjectPage_Xinyi}
``{Project Github page}.'' \url{https://github.com/xinyiW915/quality-energy}.

\bibitem{ffmpeg}
``{FFMPEG}.'' \url{https://www.ffmpeg.org}.

\bibitem{j:Ohm}
J.~R. Ohm, G.~J. Sullivan, H.~Schwarz, T.~K. Tan, and T.~Wiegand, ``Comparison
  of the coding efficiency of video coding standard - including {H}igh
  {E}fficiency {V}ideo {C}oding ({HEVC}),'' {\em IEEE Transactions on Circuits
  and Systems for Video Technology}, vol.~22, no.~12, pp.~1669--1684, 2012.

\bibitem{b:Wien}
M.~Wien, {\em {High Efficiency Video Coding: Coding Tools and Specification}}.
\newblock Springer, 2015.

\end{thebibliography}

\end{document}